\documentclass[pt12]{JHEP3} 

\JHEPspecialurl{http://jhep.sissa.it/JOURNAL/JHEP3.tar.gz}

\usepackage{epsfig,multicol}

\usepackage{amsmath, amssymb,graphicx, epic, eepic}

  \newcommand{\m}{\mu}

\newcommand{\RR}{\mathbb{R}}
\newcommand{\CC}{\mathbb{C}}
\newcommand{\ZZ}{\mathbb{Z}}

 \newcommand{\N}{{\cal N}}

\title{Towards the deconstruction of M-theory}

\author{Ingo Kirsch
                and Dan Oprisa\\
        Institut f\"ur Physik,\\ Humboldt-Universit\"at zu Berlin,\\
Newtonstra{\ss}e 15,\\ D-12489 Berlin,\\ Germany\\
        E-mail: \email{ik@physik.hu-berlin.de},
                \email{oprisa@physik.hu-berlin.de}}

\received{***, 2003}            
\revised{***, 2003}
\accepted{***, 2003}            

\preprint{\hepth{0307180}\\ HU-EP-03/40}                                 

\abstract{We argue that there is an equivalence of M-theory on $T^3 \times
  A_{N-1}$ with a four-dimensional non-supersymmetric quiver gauge theory on
  the Higgs branch.  The quiver theory in question has gauge group
  $SU(N)^{N_4N_6N_8}$ and is considered in a strong coupling and large
  $N_{4,6,8}$ limit.  We provide field- and string-theoretical evidence for
  the equivalence making use of the deconstruction technique. In particular,
  we find wrapped M2-branes in the mass spectrum of the quiver theory at low
  energies.}


\dedicated{}

\begin{document}

\section{Introduction}

Deconstruction has found many applications in both phenomenology and string
theory since its first use to generate extra dimensions in theories with
internal gauge symmetries \cite{Arkani-Hamed2001, Hill:2000mu}
(for early work on this subject, see \cite{Georgi:au, Halpern:1975yj}).
Non-renormalizable theories in $D>4$ are ill-defined above a certain cut-off
at which they become strongly coupled due to a coupling constant with negative
dimension.  At high energies these theories require an ultraviolet completion
which can frequently be provided by the deconstruction technique.  This UV
completion is generically a {\em quiver} theory \cite{Douglas:1996sw}
characterized by a discrete theory space, the {\em moose} or {\em quiver}
diagram, which represents the field content of the theory by a lattice of
sites and links. In a certain low-energy limit the quiver theory develops one
or more extra dimensions and reproduces the higher-dimensional
non-renormalizable theory. A peculiarity of the deconstruction approach is
that the ultraviolet theory has less dimensions than the infrared theory.
This is different from compactified theories which reveal their
higher-dimensional behaviour at energies above the inverse radius of
the compactified dimension.

In some instances a continuum limit of the quiver theory exists such that the
theory is higher-dimensional to arbitrary high energies.  In the known
examples the continuum theory is a theory which has been predicted to exist
by string theory arguments, but which lacks a conventional local Lagrangian
description. This has led to the deconstruction of the six-dimensional $(2,
0)$ theory and $(1, 1)$ little string theory \cite{Motl} (see also
\cite{Csaki:2002fy}) which describe the decoupling limit of multiple parallel
M5- and NS5-branes.
 Obstructions to finding a conventional local Lagrangian description of
these theories arise due to difficulties in constructing a non-abelian
chiral two-form and/or non-locality. In \cite{Motl} both theories were
deconstructed from four-dimensional quiver theories describing D3-branes at
orbifolds of the type $\CC^3/\ZZ_{N}$ and $\CC^3/\ZZ_{N_5} \times \ZZ_{N_6}$.
Another interesting theory without a Lagrangian description is that of
intersecting M5-branes.  The deconstruction of this theory
\cite{Constable,Constable:2002fb} was accomplished by taking a large $N$ limit
of the theory describing intersecting D3-branes~\cite{Constable:2002xt} at an
orbifold $\CC^2/\ZZ_N$.  Tensionless strings localized at the intersection
were identified in the spectrum of the corresponding $(4, 0)$ defect conformal
field theory.  For further applications of
deconstruction in string theory see Refs.~\cite{Rothstein:2001tu,
  Mukhi:2002ck, Alishahiha:2002jj, Giedt:2003xr, Brax, Poppitz:2003uz,
  Dai:2003ak, Iqbal}.

Recently, the deconstruction of extra dimensions in theories with gravity was
studied in \cite{Arkani-Hamed:2003vb, Arkani-Hamed:2002sp, Schwartz:2003vj,
  Jejjala, Jejjala:2003qg}.  The basic idea is to consider
\mbox{$d+1$}-dimensional \mbox{Einstein} gravity as the low-energy effective
theory of a $d$-dimen\-sional gravitational theory with a discrete theory
space.  The continuum physics of the \mbox{$d+1$}-dimensional gravitational
theory can be reproduced correctly at energies parametrically higher than the
compactification scale. However, a peculiar UV/IR connection was found
forbidding the deconstruction all the way up to the $d+1$-dimensional Planck
scale.

An open problem along these lines is the deconstruction of M-theory itself.
In~\cite{Motl} it was proposed to deconstruct M-theory on an $A_{N-1}$
singularity from a particular $(1, 0)$ little string theory (LST). This LST is
defined as the decoupling limit of $N$ NS5-branes at an orbifold singularity
of the type $\CC^2/\ZZ_{k}$. A seventh dimension arises on the
Higgs branch of this theory. In a continuum ($k\rightarrow \infty$) limit one
expects to obtain a seven-dimensional gauge theory together with its UV
completion.  Exploiting string dualities for $k\rightarrow \infty$, it was
shown~\cite{Motl} that the stack of NS5-branes maps to M-theory on $A_{N-1}$,
which is a UV completion of the seven-dimensional gauge theory.

A direct deconstruction seems to be impossible due to the obstructions to
finding a (conventional) Lagrangian description for LST.  Alternatively, one
could first deconstruct the NS5-brane theory out of the D3-brane theory at
$\CC^3/\ZZ_{N_5} \times \ZZ_{N_6}$ \cite{Motl}. One obtains a lattice action
for LST which could in principle be orbifolded again by projecting out degrees
of freedom which are not invariant under the (second) orbifold $\CC^2/\ZZ_k$.
Here one faces the problem that it is difficult, if not impossible, to find an
$SU(2)$ \mbox{R-symmetry} inside the lattice action into which the $\ZZ_k$
orbifold action can be embedded. Note that the $SO(4)$ \mbox{R-symmetry} of
the NS5-brane theory is only recovered in the continuum limit.  It is
therefore not obvious how to further discretize the latticized LST action.

In this paper we apply the deconstruction method to M-theory following a
slightly different approach. We deconstruct M-theory directly from a
four-dimensional non-super\-symmetric quiver gauge theory with gauge group
$SU(N)^{N_4N_6N_8}$ and $N_{4,6,8}$ three positive integers. The corresponding
orbifold realization is given by a stack of D3-branes in type IIB string
theory placed at the origin of $\CC^3/\Gamma$, where the orbifold group
$\Gamma$ is the product of three cyclic groups $\ZZ_{N_4} \times \ZZ_{N_6}
\times \ZZ_{N_8}$. These groups generate three circular orbits in the
directions 468.  The quiver diagram is a three-dimensional body-centred cubic
lattice. At a certain point in the moduli space, each of the $\ZZ_{N}$ factors
generates a circular discretized extra dimension.  In an appropriate $N_{4,6,8}
\rightarrow \infty$ limit, the extra dimensions become continuous, such that
the theory appears to be seven-dimensional on the Higgs branch.  There is
however a peculiarity in this deconstruction which suggests that the strongly
coupled Higgs branch theory is actually an eleven-dimensional gravitational
theory: The deconstructed seven-dimensional gauge theory has a UV completion in terms of
M-theory on $A_{N-1}$.  In the brane realization of the present
deconstruction, M-theory on $A_{N-1}$ arises naturally in the continuum limit.

The Higgs branch of the quiver theory corresponds to the decoupling limit of
\mbox{D3-branes} a finite distance away from the orbifold singularity. In the
limit which we will consider the D3-branes probe an approximate \mbox{$\RR^3
  \times T^3$} geometry.  The generation of three extra dimensions along the
Higgs branch corresponds to T-dualizing along the three circular dimensions of
the three-torus $T^3$, giving D6-branes wrapped on $T^3$. The
seven-dimensional gauge theory living on the D6-branes does not decouple from
the bulk degrees of freedom, such that the deconstructed theory is not just a
seven-dimensional gauge theory. Due to a strong type IIA string coupling $g_s$
a better description is obtained by lifting to M-theory. The D6-branes in type
IIA string theory lift to M-theory on an $A_{N-1}$ singularity.  This suggests
the equivalence of M-theory on $A_{N-1}$ with the continuum limit of the
present quiver theory.

The equivalence is also supported by the following properties of the quiver
theory on the Higgs branch. We find Kaluza-Klein states in the spectrum of
massive gauge bosons which are responsible for the generation of three extra
dimensions. Since in M-theory on $A_{N-1}$ the gauge theory is localized at
the singularity which is a seven-dimensional submanifold, we can see three of
the seven extra dimensions in the gauge boson spectrum.  Moreover, we identify
states in the spectrum of massive dyons which are identical to M2-branes
wrapping two of the three compact extra dimensions. 

The organization of this paper is as follows. In section 2 we will discuss the
quiver theory and its Higgs branch. We will find a spectrum of massive gauge
bosons and dyons indicating three extra dimensions and wrapped M2-branes.  In
section 3 we give string-theoretical evidence for the equivalence of the
continuum limit of the quiver theory with M-theory on $A_{N-1}$.  In section~4
we conclude and discuss open problems such as the relation to matrix models.

\section{Deconstruction of M-theory on $T^3 \times A_{N-1}$}

In this section we discuss the Higgs branch of the quiver theory, from which
we deconstruct M-theory. The quiver theory is a four-dimensional
non-super\-symmetric field theory with gauge group $SU(N)^{N_4N_6N_8}$. This
theory describes the decoupling limit of $NN_4N_6N_8$ \mbox{D3-branes} in type
IIB string theory placed at a $\CC^3/\Gamma$ orbifold singularity with $\Gamma
\equiv \ZZ_{N_4} \times \ZZ_{N_6} \times \ZZ_{N_8}$.  We argue that M-theory
on an $A_{N-1}$ singularity is described by the orbifold model on the Higgs
branch in a certain large $N_{4,6,8}$ and strong coupling limit.

\subsection{Non-supersymmetric orbifolds}

Before we discuss this particular product orbifold, let us briefly review
non-supersymmetric orbifold models arising from $N$ D3-branes at
$\CC^3/\Gamma$. In order to break $\N=4$ supersymmetry of the parent theory
down to $\N=0$, we choose $\Gamma$ to be a finite subgroup of the transverse
$SU(4)$ isometry group. If \mbox{$\Gamma \subset \!\!\!\!\!\!\!~/ \,\,
  SU(3)\subset SU(4)$} then no supersymmetry is preserved.  The
orbifold action is embedded into the gauge group $U(N \vert \Gamma \vert)$.

The invariant components of the gauge field satisfy
\begin{align}
  A_\mu = g(\xi) A_\mu g(\xi)^{-1} \,,      \label{transf1}
\end{align}
where $g(\xi)$ is the regular representation of the generator $\xi$ of
$\Gamma$. The gauge field $A_\m$ is a matrix in the adjoint representation of
$U(N \vert \Gamma \vert)$. Since only block-diagonal matrices $A_\m$ survive
the projection, the gauge group is broken down to $U(N)^{\vert \Gamma \vert}$.

The four Weyl fermions of $\N=4$ super Yang-Mills theory transform in the {\bf
  4} of $SU(4)$. Those fermions which are invariant under the orbifold
   must satisfy
\begin{align}
  \psi^i = \xi^{a_i} g(\xi) \psi^i g(\xi)^{-1} \,,
\end{align}
where $i=1,...,4$ and
\begin{align}
a_1+a_2+a_3+a_4 \equiv 0 \!\! \mod \vert \Gamma \vert \,. \label{condition}
\end{align}
The complex scalars $\phi^i$, $i=1,2,3$, transform in the {\bf 6} of
$SU(4)$, which can be obtained from the anti-symmetric tensor product
of two {\bf 4}'s. Invariant scalars fulfill the conditions
\begin{align}
  \phi^i = \xi^{b_i} g(\xi) \phi^i g(\xi)^{-1} \,,     \label{transf3}
\end{align}
where $b_1 = a_2 + a_3$, $b_2 = a_3 + a_1$, and $b_3 = a_1 + a_2$. A simple
example of a non-supersymmetric orbifold is $\CC/\ZZ_{N'}$ defined by the
vectors
\begin{align}
a_i=(-1,1,1,-1)\,, \quad b_i=(2,0,0) \,.  \label{nonsusy}
\end{align}
Since $b_2=b_3=0$ the orbifold acts only in one of
the three complex planes transverse to the D3-branes.
The regular representation of the
generator $\xi=e^{2\pi i/{N'}}$ of the group $\ZZ_{N'}$ is given by
$g(\xi)={\rm diag}(I,\xi I, \xi^2 I, ..., \xi^{N'-1}I)$.

\subsection{D3-branes at $\CC^3/\ZZ_{N_4} \times \ZZ_{N_6} \times 
  \ZZ_{N_8}$: orbifold \mbox{realization} of a non-super\-sym\-metric quiver
  theory}

We now turn to the case where $N N_4 N_6 N_8$ D3-branes are placed at an
orbifold $\CC^3/\Gamma$ with \linebreak \mbox{$\Gamma \equiv \ZZ_{N_4} \times
  \ZZ_{N_6} \times \ZZ_{N_8}$}. The orbifold action on the complex coordinates
$z_i=(h, v, n)$ parametrizing $\CC^3$ is given by
\begin{align}  \label{orbaction}
h \rightarrow \xi_4^{\,2} h\,,\qquad
v \rightarrow \xi_6^{\,2} v\,,\qquad
n \rightarrow \xi_8^{\,2} n\,,
\end{align}  
where the generators of the groups $\ZZ_{N_a}$ are defined by $\xi_a =\exp(2\pi
i/N_a)$, $a=4, 6, 8$. Each of the factors $\ZZ_{N_a}$ ($a=4,6,8$) acts on one
of the three complex planes transverse to the stack of
D3-branes. The orbifold action can be embedded into the subgroup $U(1)^3$
of the rotational group $SO(6)$.

The fields of the quiver theory descend from the $\N=4, d=4$ vector multiplet
of the parent super Yang-Mills theory with gauge group $U(N N_4N_6N_8)$.
We project out degrees of freedom which are not
invariant under the orbifold group. The action of the product orbifold on
the gauge field $A_\mu$, the three scalars $\phi^i=(h,v,n)$, and
the four spinors $\psi^i=(\psi^h, \psi^v, \psi^n, \lambda)$ is given by
\begin{align}
  A_\mu &\rightarrow g(\xi) A_\mu g(\xi)^{-1} \,, \\
  \phi^i &\rightarrow \xi^{b^{(4)}_i}_4 \xi^{b^{(6)}_i}_6 \xi^{b^{(8)}_i}_8
            g(\xi) \phi^i g(\xi)^{-1} \,, \\
  \psi^i &\rightarrow \xi^{a^{(4)}_i}_4 \xi^{a^{(6)}_i}_6 \xi^{a^{(8)}_i}_8
            g(\xi) \psi^i g(\xi)^{-1} \,,
\end{align}
where $g(\xi)=g(\xi_4) \otimes g(\xi_6) \otimes g(\xi_8)$ is the
regular representation of the generator $\xi=\xi_4 \xi_6 \xi_8$ of $\Gamma$.
These transformation rules extend the invariance conditions given by
Eqns.~(\ref{transf1})-(\ref{transf3}).
For the vectors $a_i$ and $b_i$ we choose:
\begin{align}
 \ZZ_{N_4}:\quad a^{(4)}_i&=(-1,1,1,-1) \,,\quad b^{(4)}_i=(2,0,0) \,,
 \label{tuple1}\\
 \ZZ_{N_6}:\quad a^{(6)}_i&=(1,-1,1,-1) \,,\quad b^{(6)}_i=(0,2,0) \,,\\
 \ZZ_{N_8}:\quad a^{(8)}_i&=(1,1,-1,-1) \,,\quad b^{(8)}_i=(0,0,2) \,.
 \label{tuple3}
\end{align}

Each of the three pairs of vectors ($a_i, b_i$) gives rise to an orbifold
$\CC/\ZZ_N$ of the type given by Eq.~(\ref{nonsusy}). Together they define the
non-supersymmetric orbifold $\CC^3/\ZZ_{N_4} \times \ZZ_{N_6} \times
\ZZ_{N_8}$. The vectors $b_i$ determine the action (\ref{orbaction}) on the
coordinates $z_i=(h, v, n)$ via 
\begin{align}
z_i \rightarrow \xi^{b^{(4)}_i}_4 \xi^{b^{(6)}_i}_6 \xi^{b^{(8)}_i}_8 z_i\,.
\end{align}
The vectors $a_i$ give the corresponding action on the four fermions. The
invariant fermions $\psi^m_{i,j,k}$ transform under the gauge group
\begin{align}
  SU(N)^{N_4 N_6 N_8}
\end{align}
as $({\rm N}_{i,j,k}, \overline{\rm N}_{i\pm a^{(4)}_m, j\pm a^{(6)}_m, k \pm
  a^{(8)}_m})$, where N$_{i,j,k}$ ($\overline{\rm N}_{i',j',k'}$) denotes the
(anti-)funda\-mental representation of the gauge group $SU(N)_{i,j,k}$
($SU(N)_{i',j',k'}$). The invariant scalars $\phi^m_{i,j,k}$ transform as
$({\rm N}_{i,j,k}, \overline{\rm N}_{i\pm b^{(4)}_m, j\pm b^{(6)}_m, k\pm
  b^{(8)}_m})$. We summarize the field content of our quiver theory in the
following table:

\begin{table}[!h] \label{matter}
\begin{center}
\begin{tabular}{|l|l||l|l|} 
\hline
  field & representation & field & representation \\
\hline
  $h_{i,j,k}$ & $({\rm N}_{i,j,k}, \overline{\rm N}_{i+2,j,k})$ &
  $\psi^h_{i,j,k}$  & $({\rm N}_{i,j,k}, \overline{\rm N}_{i+1,j-1,k-1})$\\
  $v_{i,j,k}$ & $({\rm N}_{i,j,k}, \overline{\rm N}_{i,j+2,k})$ &
  $\psi^v_{i,j,k}$  & $({\rm N}_{i,j,k}, \overline{\rm N}_{i-1,j+1,k-1})$\\
  $n_{i,j,k}$ & $({\rm N}_{i,j,k}, \overline{\rm N}_{i,j,k+2})$ &
  $\psi^n_{i,j,k}$  & $({\rm N}_{i,j,k}, \overline{\rm N}_{i-1,j-1,k+1})$\\
  $A^\mu_{i,j,k}$ & {\rm adjoint} &
  $\lambda_{i,j,k}$ & $({\rm N}_{i,j,k}, \overline{\rm N}_{i+1,j+1,k+1})$\\
 \hline
\end{tabular}
\caption{Fields in the quiver theory and their transformation behaviour
under the gauge group $SU(N)^{N_4 N_6 N_8}$.} \label{tab1}
\end{center}
\end{table}

\FIGURE{\epsfig{file=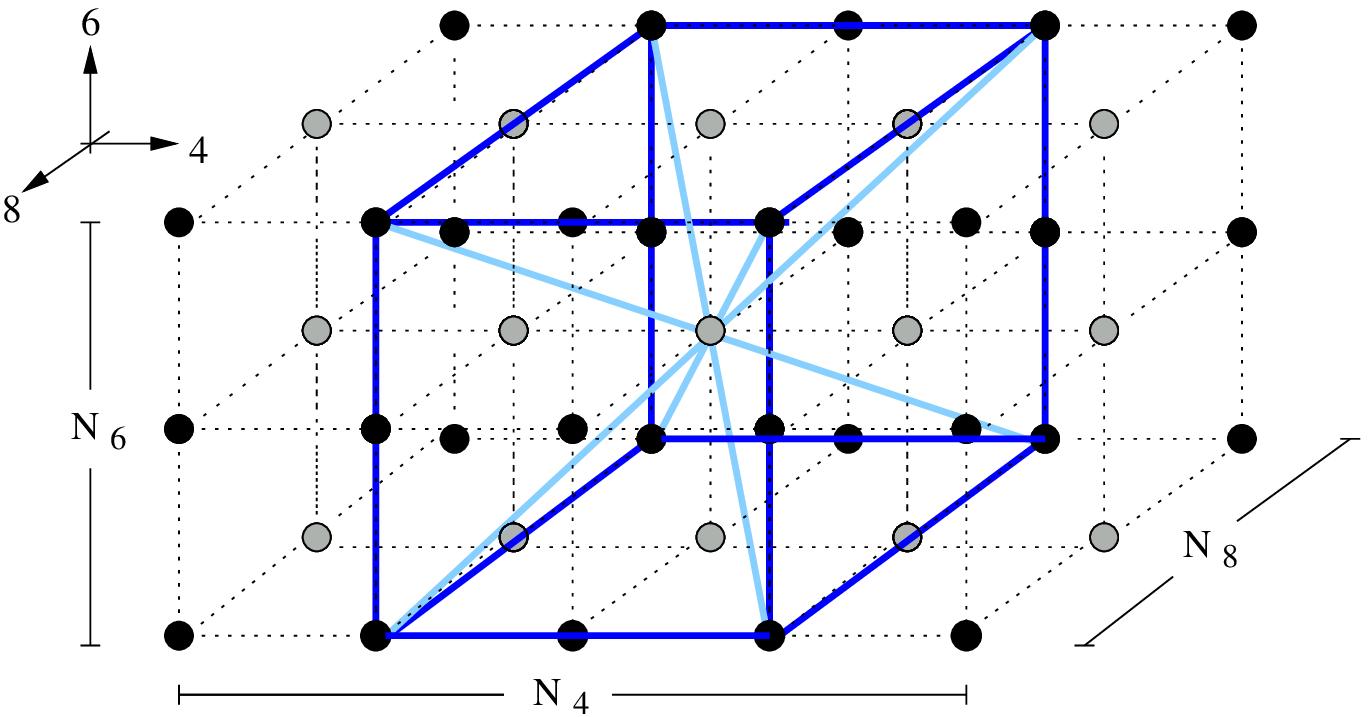,scale=.75} 
        \caption[Example of figure]{Theory space for the $\ZZ_{N_4} \times \ZZ_{N_6} \times
  \ZZ_{N_8}$ quiver theory. Dark and light (blue) lines in the basic cell
  represent bosonic and fermionic bifundamentals. Dotted lines are not
  physical and are just to guide the eye.}
        \label{box3}}

The theory space is a three-dimensional lattice with $N_4N_6N_8$ sites which
discretizes a three-dimensional torus $T^3$ as shown in Fig.~\ref{box3}.  Each
site represents one of the gauge groups $SU(N)_{i,j,k}$ and its associated
gauge boson~$A^\mu_{i,j,k}$. Link fields start at a site $i,j,k$, where they
transform in the fundamental representation ${\rm N}_{i,j,k}$, and end at a
site $i',j',k'$, where they transform in the antifundamental representation
$\overline{\rm N}_{i',j',k'}$. Fig.~{\ref{box3} shows the unit cell of the
  lattice spanned by the link fields. The bosonic bifundamentals $h_{i,j,k}$,
  $v_{i,j,k}$, $n_{i,j,k}$ (dark lines) form the edges of the unit cell,
  while the fermionic bifundamentals $\psi^h_{i,j,k}, \psi^v_{i,j,k},
  \psi^n_{i,j,k}, \lambda_{i,j,k}$ (light lines) connect the corners
  with the centre of the cell. Translating the unit cell in the lattice
we obtain a  body-centred cubic lattice which is invariant under the 48 element
octahedral symmetry group~$O_h$. Such bcc lattices were also studied in the
context of four-dimensional $\N=4$ super Yang-Mills theory on a
three-dimensional lattice \cite{Kaplan}.

Let us now construct the Lagrangian for our orbifold model which consists of
three parts,
\begin{align}
  {\cal L}={\cal L}_{\rm kin} + {\cal L}_{\rm Yuk} + {\cal L}_{\rm quartic}
  \,.  \label{action}
\end{align}
It follows from the four-dimensional $\N=4$ super Yang-Mills theory with gauge
group $U(NN_4N_6N_8)$ upon projecting out degrees of freedom which are not
invariant under the orbifold group. The kinetic terms have the form
\begin{align}
  {\cal L}_{\rm kin} \supset \frac{1}{2}{\rm Tr} (D_\mu
    \varphi_{i',j',k'})^\dagger D^\mu \varphi_{i,j,k} \,,
\end{align}
where the field $\varphi_{i,j,k}$ is one of the seven bifundamentals
transforming in the $({\rm N}_{i,j,k}, \overline{\rm
  N}_{i',j',k'})$ as listed in Tab.~\ref{tab1}. The covariant derivative of
$\varphi_{i,j,k}$ is defined by
\begin{align}
  D_\mu \varphi_{i,j,k} = \partial_\mu \varphi_{i,j,k} - i g A^{i,j,k}_\mu
  \varphi_{i,j,k} + i g \varphi_{i,j,k} A^{i',j',k'}_\mu \,.
\end{align}

\FIGURE{\epsfig{file=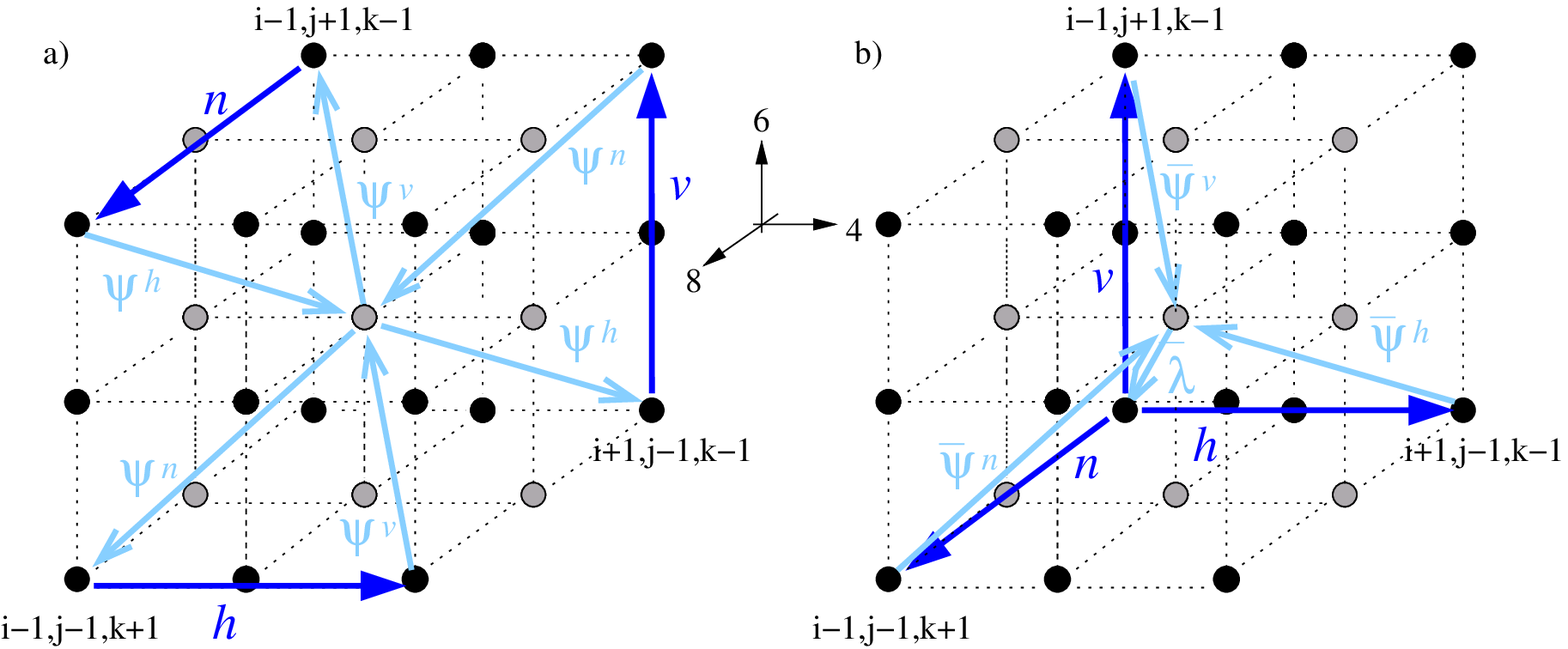,scale=.75} 
        \caption[Example of figure]{Oriented link fields in the basic cell. The site in the
          centre of the cell has labels $i, j, k$.  The triangles represent
          possible Yukawa couplings in the quiver action.}
        \label{box}}
We now consider the Yukawa and quartic scalar interactions ${\cal L}_{\rm
  Yuk}$ and ${\cal L}_{\rm quartic}$.  Fig.~\ref{box} shows six of twelve
possible triangles inside the basic cell.  These triangles consist of two
fermionic and one bosonic arrow and correspond to Yukawa couplings in the
quiver theory.  Each of the twelve triangles leads to a Yukawa term in the
action.  

For the Lagrangian ${\cal L}_{\rm Yuk}$ we thus find
\begin{align}
{\cal L}_{\rm Yuk}={\cal L}^1_{\rm Yuk}+{\cal L}^2_{\rm Yuk} \,,\label{Yukawa}
\end{align}
with
\begin{align}
  \!\!\!\!\!\!{\cal L}^1_{\rm Yuk}&= i\sqrt{2} g {\,\rm Tr}(\psi^v_{i,j,k}
  n_{i-1,j+1,k-1} \psi^h_{i-1,j+1,k+1}
  - \psi^n_{i,j,k} v_{i-1,j-1,k+1} \psi^h_{i-1,j+1,k+1} \nonumber\\
  &\,+ \psi^n_{i,j,k} h_{i-1,j-1,k+1} \psi^v_{i+1,j-1,k+1}
  - \psi^h_{i,j,k} n_{i+1,j-1,k-1} \psi^v_{i+1,j-1,k+1} \\
  &\,+ \psi^h_{i,j,k} v_{i+1,j-1,k-1} \psi^v_{i+1,j+1,k-1} - \psi^v_{i,j,k}
  h_{i-1,j+1,k-1} \psi^n_{i+1,j+1,k-1} + c.c.) \nonumber
\end{align}
and 
\begin{align}
  \!\!\!\!\!\!{\cal L}^2_{\rm Yuk}&= i\sqrt{2} g {\,\rm Tr}(\bar\lambda_{i,j,k} h_{i-1,j-1,k-1} \bar\psi^h_{i+1,j-1,k-1}
   - \bar \psi^h_{i,j,k} h_{i-1,j+1,k+1}  \bar \lambda_{i+1,j+1,k+1}\nonumber\\
  &+ \bar\lambda_{i,j,k} v_{i-1,j-1,k-1} \bar \psi^v_{i-1,j+1,k-1} 
   - \bar \psi^v_{i,j,k} v_{i+1,j-1,k+1} \bar\lambda_{i+1,j+1,k+1}  \\
  &+ \bar\lambda_{i,j,k} n_{i-1,j-1,k-1} \bar \psi^n_{i-1,j-1,k+1} 
   - \bar \psi^n_{i,j,k} n_{i+1,j+1,k-1} \bar\lambda_{i+1,j+1,k+1} + c.c.)
   \nonumber \,,
\end{align} 
where summation over $i,j,k$ is understood. The terms with a positive sign in
${\cal L}^{1}_{\rm Yuk}$(${\cal L}^{2}_{\rm Yuk}$) correspond to the triangles
in Fig.~\ref{box}a(b) (those with negative sign are not shown).  The Yukawa
couplings in ${\cal L}^{1}_{\rm Yuk}$ descend from the $\N=4$ superpotential
$[H,V]N$, while those in ${\cal L}^{2}_{\rm Yuk}$ come from the K\"ahler
potential.

 Quite analogously, squares in the quiver diagram represent quartic
scalar terms which are given by\footnote{There are some additional terms
  contributing to ${\cal L}_{\rm quartic}$: For brevity we did not list terms
  corresponding to degenerate squares coming from D-terms like $D\bar hh+D\bar
  vv+ D\bar nn$.}
\begin{align}
\!\!\!\!\!\!{\cal L}_{\rm quartic}  
&=g^2 {\,\rm Tr}(n_{i,j,k}h_{i,j,k+2}\bar{n}_{i+2,j,k+2}\bar{h}_{i+2,j,k}
 - h_{i,j,k}n_{i+2,j,k}\bar{h}_{i+2,j,k+2}\bar{n}_{i,j,k+2}\nonumber\\
&+h_{i,j,k}v_{i+2,j,k}\bar{h}_{i+2,j+2,k}\bar{v}_{i,j+2,k}
 -v_{i,j,k}h_{i,j+2,k}\bar{v}_{i+2,j+2,k}\bar{h}_{i+2,j,k}\\
&+v_{i,j,k}n_{i,j+2,k}\bar{v}_{i,j+2,k+2}\bar{n}_{i,j,k+2}
 -n_{i,j,k}v_{i,j,k+2}\bar{n}_{i,j+2,k+2}\bar{v}_{i,j+2,k})\nonumber
  \,. \label{scalarterms}
\end{align} 

This model belongs to the class of conformal non-supersymmetric orbifold
models studied in~\cite{Lawrence, Bershadsky}. In these models it is assumed
that the orbifold group $\Gamma \subset SU(4)$ acts solely on the transverse
space $\CC^3$ of $M$ parallel D3-branes. Kachru and Silverstein \cite{Kachru}
noticed that the orbifold group acts only on the $S^5$ factor of the near
horizon geometry \mbox{$AdS_5 \times S^5$}. In the AdS/CFT correspondence the
isometry group of the $AdS_5$ space is identified with the conformal group of
the field theory on its boundary.  This implies classical conformal invariance
of the worldvolume theory on the D3-branes. In \cite{Lawrence} it was shown
that if $M$ is finite and the regular representation of $\Gamma$ is chosen,
the one-loop beta functions for the gauge couplings vanish in these theories.
In the large $M$ limit one can even prove the vanishing of the beta functions
to all orders in perturbation theory \cite{Bershadsky}. This holds for the
present quiver theory which we consider in the limit $N_{4,6,8} \rightarrow
\infty$ such that $M=N\,N_4N_6N_8 \rightarrow \infty$.\footnote{ The basic
  requirement for vanishing beta function, Eq.~(2.7) in \cite{Bershadsky}, is
  satisfied: Let $\gamma^a \in \Gamma \equiv \ZZ_{N_4} \times \ZZ_{N_6} \times
  \ZZ_{N_8}$ and $\gamma^a_{4,6,8} \in \ZZ_{N_{4,6,8}}$. Then
\begin{align} 
  {\rm tr}(\gamma^a)&=(1+\gamma_4^a+...+(\gamma_4^a)^{N_4-1})
  (1+...+(\gamma_6^a)^{N_6-1})(1+...+(\gamma_8^a)^{N_8-1})
  N \nonumber\\
  &= \left\lbrace \begin{tabular}{l} $N_4N_6N_8 N$ if $\gamma^1=1$\\
      $0$ $\forall \gamma^a$, $a \neq 1$ \end{tabular} \right. .  \nonumber
\end{align}
}${}^{,3}$ Our non-supersymmetric quiver theory
has therefore quantum conformal invariance. As discussed in detail in
\cite{Motl} conformal invariance guarantees that the quiver theory remains in
the Higgs phase even at strong coupling.

A related question to that of conformal invariance is the stability of
the moduli space. Since the theory is not supersymmetric the potential
for the scalars is not necessarily protected against quantum corrections.
This would change the moduli space of the classical theory. However, as
shown in \cite{Bershadsky} all Feynman diagrams in the quiver theory
are the same as in the $\N=4$ parent $SU(M)$ gauge theory up to possible $\frac{1}{M}$
corrections. In the large $M$ limit these corrections are suppressed and
the potential remains unchanged.\footnote{Note that it is not necessary
to send $N \rightarrow \infty$ in order to take $M=N N_4N_6N_8 \rightarrow
\infty$.} Although this points to a stable moduli space, we might still
have troubles with divergencies coming from the twisted closed string sector.

The question of stability of our model is highly non-trivial.  In contrast to
supersymmetric orbifold models there are {\em closed string tachyons} in the
twisted sector of non-supersymmetric orbifolds of the type $\CC/\ZZ_N$.  The
tachyon condensation leads to the decay of the orbifold as studied in
\cite{Adams}.  The initial effect of the tachyons is to smooth out the
orbifold singularity.  An RG flow is initiated by the tachyon and the orbifold
decays to flat space. If the initial state has been appropriately fine-tuned,
the orbifold decay can take place in a series of transitions $\CC/\ZZ_N
\rightarrow \CC/\ZZ_{N-2}$.  For finite $N$ the orbifold becomes flat in a
finite time. However, in the limit $N \rightarrow \infty$ the orbifold does
not decay in a {\em finite} amount of time, since the orbifold goes through
only finitely many transitions $\CC/\ZZ_N \rightarrow \CC/\ZZ_{N-2}$.  The
orbifold may however decay faster, e.g.~via transitions $\CC/\ZZ_N \rightarrow
\CC/\ZZ_{N-M}$ ($M>2$). If the quotient $M/N$ vanishes in the large $N$ limit
the orbifold remains stable.  In other words, the question is whether the
flattening of spacetime induced by the tachyon condensation outweighs the
extreme curvature at the singularity.  We believe it does not and presume that
our model is stable in the large $N_{4,6,8}$ limit. However, this issue
deserves some further investigation.

Another essential feature in the deconstruction of M-theory is S-duality of
the orbifold model.  In \cite{Lawrence} it was argued that in a conformal
(non-super\-sym\-metric) quiver theory the complex moduli $\tau_i$ are
inherited from the coupling $\tau_{\rm par}$ of the $\N=4$ parent theory
(recall $\tau_i=\tau_{\rm par}/\vert \Gamma\vert$). In the present quiver
theory the $N_4N_6N_8$ gauge couplings $\tau_{i,j,k}$ associated with the
gauge groups $SU(N)_{i,j,k}$ are all the same and related to $\tau_{\rm par}$
by
\begin{align}
 \tau \equiv \tau_{i,j,k}=\frac{\tau_{\rm par}}{N_4N_6N_8} \,.
\end{align}
The strong-weak duality $g_{\rm par} \rightarrow 1/g_{\rm par}$ thus
amounts to an $SL(2,\ZZ)$ S-duality \linebreak \mbox{$g \rightarrow N_4 N_6 N_8/g$} in
the quiver theory.

\subsection{Generation of three compact extra dimensions
  in the low-energy effective field theory}

We now show by studying the mass spectrum of the gauge bosons that the quiver
theory generates three circular extra dimensions at low energies.  On the
Higgs branch of the theory the scalar bifundamentals have diagonal expectation
values,
\begin{align}
    \langle h_{i,j,k} \rangle=v_4\,, \qquad \langle v_{i,j,k} \rangle=v_6\,,
    \qquad \langle n_{i,j,k} \rangle=v_8 \,,
\end{align}
independent of $i,j,k$. These condensates break the gauge group
$SU(N)^{N_4N_6N_8}$ down to the diagonal subgroup $SU(N)$.  Upon substituting
the {\em vevs} $v_4, v_6, v_8$, the scalar kinetic terms inside ${\cal L}_{\rm
  kin}$ give rise to mass terms for the gauge bosons,
\begin{align}
  {\rm Tr} \vert D^\m h_{i,j,k} \vert^2 
  &=  g^2v^2_4 (A_\mu^{i,j,k} - A_\mu^{i+2,j,k})^2  
   \equiv A_\mu^{i,j,k} ({\cal M}_4)^2_{ii'} \delta_{jj'} \delta_{kk'} 
   A^\mu_{i',j',k'}\nonumber\\
 {\rm Tr} \vert D^\m v_{i,j,k} \vert^2
  &=  g^2v^2_6 (A_\mu^{i,j,k} - A_\mu^{i,j+2,k})^2 
   \equiv A_\mu^{i,j,k} \delta_{ii'} ({\cal M}_6)^2_{jj'} \delta_{kk'}
   A^\mu_{i',j',k'}\\
 {\rm Tr} \vert D^\m n_{i,j,k} \vert^2
  &=  g^2v^2_8 (A_\mu^{i,j,k} - A_\mu^{i,j,k+2})^2 
   \equiv A_\mu^{i,j,k} \delta_{ii'} \delta_{jj'} ({\cal M}_8)^2_{kk'} 
   A^\mu_{i',j',k'} \,.\nonumber
\end{align}
The matrices ${\cal M}_{4,6,8}$ have entries $2$ on the diagonal and $-1$ on
the second off-diagonal.  As in \cite{Arkani-Hamed2001} diagonalization
of the mass matrices ${\cal M}_{4,6,8}$ yields the mass eigenvalues
\begin{align}
m^k_{4,6,8} = 2 g v_{4,6,8} \sin \frac{2 \pi k}{N_{4,6,8}} 
\approx 2 g v_{4,6,8}\frac{2 \pi k}{N_{4,6,8}} \quad{\rm for\,\,} 
k \ll N_{4,6,8}\,. 
\end{align}
For small enough $k$, this approximates the Kaluza-Klein spectrum of a
seven-dimensional gauge boson compactified on a three-torus $T^3$ with radii
$R_{4,6,8}$.  The radii $R_{4,6,8}$ are fixed by the mass scales of the
lightest KK modes ($k=1$) which are given by
\begin{align}
m_{4,6,8} 
= \frac{1}{R_{4,6,8}} \,,\label{mass}
\end{align}
with
\begin{align} \label{radii}
2 \pi R_4 = N_4 a_4 = \frac{N_4}{2gv_4} \,, \quad
2 \pi R_6 = N_6 a_6 = \frac{N_6}{2gv_6} \,,\quad
2 \pi R_8 = N_8 a_8 = \frac{N_8}{2gv_8} \,
\end{align}
and $a_{4,6,8}$ the lattice spacings. 
In principle, this  KK spectrum is not protected and could receive quantum
corrections at strong coupling. In a similar context \cite{Motl} it was argued
that such quantum corrections are proportional to $\frac{1}{N_{4,6,8}}$ and
vanish in the large $N_{4,6,8}$ limit.  Provided this is true, we explicitly
deconstruct three compact extra dimensions with radii $R_{4,6,8}$. 

The Higgs mechanism does induce masses both for the gauge bosons as well
as for the bifundamental fermions and scalars. For instance, substituting {\em
  vevs} for the scalars inside the Lagrangian ${\cal L}_{\rm Yuk}$ leads to
fermionic mass terms. Such mass terms could in principle lead to a different
mass spectrum due to the non-supersymmetric nature of our model. Following
\cite{Csaki} one can however verify that the fermionic mass spectrum is
identical to the gauge boson spectrum.  Both the bosonic as well as the
fermionic Kaluza-Klein spectra generate the same extra dimensions.

\subsection{M2-branes on the Higgs branch}

By studying the orbifold geometry we show in the next section that the Higgs
branch theory is equivalent to M-theory on an $A_{N-1}$ singular geometry. We
have seen that the Higgs branch theory contains seven-dimensional super
Yang-Mills theory with gauge group $SU(N)$.  In M-theory on $A_{N-1}$ the
seven-di\-men\-sional $SU(N)$ gauge symmetry arises from M2-branes wrapping
collapsed two-cycles at the singularity, see e.g.~\cite{Sen}. In other words,
the deconstructed 7d super Yang-Mills theory is part of M-theory on $A_{N-1}$.
However, \mbox{M-theory} contains more than just the 7d gauge theory. We have
to verify also the existence of M2- or M5-branes inside the quiver theory.

The states corresponding to M2-branes can be seen in the dyonic spectrum of
the quiver gauge theory.  The dyonic mass spectrum follows from that of the
gauge bosons by S-duality.  Substituting
\begin{align}
 g \rightarrow \frac{N_4N_6N_8}{g} \,
\end{align}
into Eq.~(\ref{mass}), we find for the lowest dyonic states the masses
\begin{align}
M_4 = 8\pi^3 R_6 R_8/g^2_7 \,,\qquad M_6 = 8\pi^3 R_8 R_4/g^2_7 \,,\qquad
M_8 = 8\pi^3 R_4 R_6/g^2_7 \,,
\end{align}
where the seven-dimensional coupling constant is $g_7^2 = a_4 a_6 a_8 g^2$.
These masses are identical to those of two-branes wrapping around
two-tori~$T^2$ inside the $T^3$. We can read off the tension of the
two-branes, {$T_{2}= 1/(2\pi)^2 g_7^2$}, which is identical to the tension of
M2-branes, $T_{\rm M2} = 1/(2\pi)^2 l_p^3$.  This gives field-theoretical
evidence that we really deconstruct M-theory.\footnote{We cannot see M5-branes
  in this way.  The theory we expect to deconstruct is \mbox{M-theory} on the
  geometry $\RR^{1,3} \times T^3 \times A_{N-1}$.  There are not enough
  compact dimensions inside this geometry which M5-branes could wrap around.}

\subsection{Summary of the field theory results}

Let us summarize the properties of the $SU(N)^{N_4N_6N_8}$ quiver theory.  We
have seen that three extra dimensions with fixed radii $R_{4,6,8}$ are
generated along its diagonal Higgs branch.  For finite lattice spacings
$a_{4,6,8}$ our four-dimensional quiver theory describes a seven-dimensional
theory with gauge coupling $g_7^2$ discretized on a three-dimensional toroidal
lattice. Seven-dimensional super Yang-Mills theory breaks down at a certain
cut-off $\Lambda_{\rm 7d}$ above which it requires a UV completion.  The
cut-off of the deconstructed theory is given by the mass of the highest KK
mode, $\Lambda=a^{-1}$ ($a=\max [a_4, a_6, a_8]$). In the continuum limit
$a_{4,6,8} \rightarrow 0$, which requires \mbox{$g\rightarrow\infty$} while
keeping the radii $R_{4,6,8}$ and the seven-dimensional gauge coupling $g_7^2$
fixed, $\Lambda$ becomes very large, $\Lambda \gg \Lambda_{\rm 7d}$.  In the
large $N_{4,6,8}$ limit we therefore expect to deconstruct not only 7d super
Yang-Mills theory but also its UV completion. This is shown schematically in
Fig.~\ref{energy}.  We show in the next section that the UV completion is
M-theory on $A_{N-1}$ with Planck length $l_p^3=g_7^2$.

\FIGURE{\epsfig{file=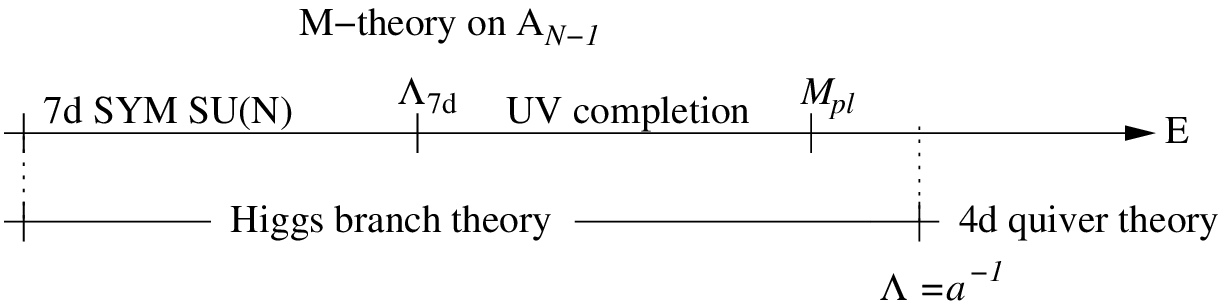,scale=1} 
        \caption{Various cut-offs in the deconstruction of M-theory.}
        \label{energy}}

M-theory on this geometry consists of two parts:~The seven-dimensional gauge
theory living on the singularity of the $A_{N-1}$ space couples to the
eleven-dimensional bulk degrees of freedom of M-theory. In the gauge boson
spectrum, we can therefore see only three of the seven string-theoretically
predicted extra dimensions.  It is not quite clear how the four dimensions of
the $A_{N-1}$ space are generated. However, we have found a spectrum of
massive dyons showing the presence of M2-branes in the deconstructed theory.
This supports our conjecture that the deconstructed theory is M-theory on
$A_{N-1}$.

\section{String-theoretical motivation for the equivalence} \label{sec25}
                
We now motivate the conjecture of the last section by a string-theoretical
analysis of the orbifold geometry. We show that supersymmetry with 16
supercharges is restored on the Higgs branch of our model, where the theory
becomes M-theory on a $T^3 \times A_{N-1}$ geometry. To this end, we consider
the behaviour of the stack of D3-branes on the Higgs branch.
                     
For this purpose, let us study the geometry of the orbifold in the vicinity
of the {D3-branes} which are located a distance $d$ away from the orbifold
singularity.  The product orbifold $\CC^3/\ZZ_{N_4} \times \ZZ_{N_6} \times
\ZZ_{N_8}$ can be parametrized by the complex coordinates 
($i=1,2,3$)~\cite{Brax}
\begin{align}
z_i = r_i \exp\left(i \frac{b^{(4)}_i}{N_4} \theta_1 + 
i \frac{b^{(6)}_i}{N_6} \theta_2 + 
i \frac{b^{(8)}_i}{N_8} \theta_3  \right) \,.
\end{align}
For orthogonal vectors $b^{(4)}_i, b^{(6)}_i, b^{(8)}_i$, the orbifold metric
$ds^2 = dz_i d\bar z_i$ \mbox{acquires} the form
\begin{align}
ds^2 = 
 d\vec r^{\,2} +   \frac{({\vec r} \cdot {\vec b}^{(4)})^2}{N_4^2} d\theta_1^2
+ \frac{({\vec r} \cdot {\vec b}^{(6)})^2 }{N_6^2} d\theta_2^2
+ \frac{({\vec r} \cdot {\vec b}^{(8)})^2 }{N_8^2} d\theta_3^2  \,,
\end{align}
where $\vec r=(r_1,r_2,r_3)$, $\vec b^{(4)}= (b^{(4)}_1,b^{(4)}_2,b^{(4)}_3)$,
etc. The orbifold has the geometry of a three-torus fibration over $\RR^3$: We
recover three circles $S^1$ parametrized by $\theta_{1,2,3} \in [0,2\pi]$.
For the particular choice of vectors $b_i$ as defined in
Eqns.~(\ref{tuple1})-(\ref{tuple3}), the circles $S^1$ have radii
$R_{S^1}= {l_s^2}/{R_{4,6,8}}$ with the radii $R_{4,6,8}$ given by
\begin{align}
R_{4}=\frac{N_{4} l_s^2}{2 d_4} \,,\qquad
R_{6}=\frac{N_{6} l_s^2}{2 d_6} \,,\qquad
R_{8}=\frac{N_{8} l_s^2}{2 d_8}\,.
\end{align}
Here the D3-branes were assumed to be located at $\vec r = (d_4,d_6,d_8)$.
Comparison with the radii defined in (\ref{radii}) yields relations between
the parameters of the quiver theory $g$, $v_{4,6,8}$, $N_{4,6,8}$ and the
string theory parameters $l_s$, $g_s$, $d_{4,6,8}=\vert z_{1,2,3} \vert$,
$N_{4,6,8}$:
\begin{align}
\frac{d_{4}}{N_{4} l_s^2} = \frac{2\pi gv_4}{N_{4}}  \,,\qquad
\frac{d_{6}}{N_{6} l_s^2} = \frac{2\pi gv_6}{N_{6}}  \,,\qquad
\frac{d_{8}}{N_{8} l_s^2} = \frac{2\pi gv_8}{N_{8}}  \,.
\end{align}
These relations show that giving {\em vevs} $v_4, v_6, v_8$ to the scalar
bifundamentals $h, v, n$ corresponds to moving the D3-branes a distance $d
=\sqrt{d_4^2+d_6^2+d_8^2}$ away from the singularity. The continuum limit $a
\rightarrow 0$ keeping $R_{4,6,8}$ fixed is obtained if we take $l_s
\rightarrow 0$, $N_{4,6,8}\rightarrow \infty$ with $g_s=g^2/N_4N_6N_8$ and
$d_{4,6,8}/ N_{4,6,8} l_s^2$ fixed.
 
The orbifold may be visualized as a product of three cones.  In the large
$N_{4,6,8}$ limit each of the cones locally degenerates into a cylinder
$\RR\times S^1$ similar as in \cite{Motl, Brax}. The orbifold geometry in the
vicinity of the D3-branes becomes approximately $\RR^3 \times T^3$ with $T^3$
a three-torus.  Note that this induces a strong {\em
  supersymmetry enhancement} in the world volume theory.  The $\N=4$ super
Yang-Mills parent theory preserves 16 supercharges.  The orbifold projection
reduces supersymmetry to $\N=0$. Now in the large $N_{4,6,8}$ limit, the
D3-branes probe the geometry $\RR^3 \times T^3$, which in contrast to the
orbifold, does not break supersymmetry. On the Higgs branch the supersymmetry
of the quiver theory is therefore enhanced again to 16 supercharges.

In the large $N_{4,6,8}$ limit the radii of the three circles $S^1$ are
sub-stringy, i.e.~$R_{S^1} \ll l_s$ if $R_{4,6,8} \gg l_s$, and the
appropriate description is obtained by T-dualizing along the three directions
of the torus~468.  Details of the T-dualization are shown in the following
table:

\begin{table}[!h]
\hspace{1.8cm}
\begin{tabular}{|c|c|l|l|} 
\hline duality
  & D3 & $l_s \rightarrow 0$ & $g_s$ fixed \\
\hline
   T in $x^4$ & D4 & $l'_s=l_s \rightarrow 0$ 
              & $g'_s=g_s R_4/l_s \rightarrow \infty$\\
   T in $x^6$ & D5 & $l''_s=l_s \rightarrow 0$
              & $g''_s=g_s R_4R_6/l^2_s \rightarrow \infty$ \\
   T in $x^8$ & D6 & $l'''_s=l_s \rightarrow 0$
              & $g'''_s=g_s R_4R_6R_8/l^3_s \rightarrow \infty$ \\
\hline
  M-theory & M-theory & $l^3_p=(l'''_s)^3 {g'''_s}$& $R_{11} =l'''_s {g'''_s}$\\
    lift  & on $A_{N-1}$ & fixed &\\
\hline
\end{tabular}
\caption{T-dualization in $x^{4,6,8}$ and lift to M-theory.}
\end{table}

We started from D3-branes in the decoupling limit $l_s \rightarrow 0$, $g_s$
fixed.  These D3-branes T-dualize to D6-branes which wrap a three-torus $T^3$
with large radii $R_{4,6,8}$.  Due to a large string coupling $g'''_s$, a more
appropriate description is obtained by lifting the stack of $N$ D6-branes to a
singular $A_{N-1}$ geometry in M-theory.  The seven-dimensional gauge theory
located on the $A_{N-1}$ singularity has gauge coupling
$g_7^2=l_p^3=(l'''_s)^3 {g'''_s}=l^2_s R_{11}$. Since $R_{11}=g_s
R_4R_6R_8/l^2_s \rightarrow \infty$ and $l_s \rightarrow 0$, we can hold both
the gauge coupling $g_7$ and the eleven-dimensional Planck length $l_p$ fixed.
The seven-dimensional gauge theory therefore does not decouple from the bulk
gravity.

To conclude, string theory arguments suggest that our strongly coupled
non-super\-symmetric quiver theory on the Higgs branch describes M-theory on
$T^3 \times A_{N-1}$ in a large $N_{4,6,8}$ limit.  Since M-theory reduces to
eleven-dimensional supergravity at low-energies, we have deconstructed  a
gravitational theory!

\section{Conclusions}

We have deconstructed M-theory on a singular space of the type 
$T^3\times A_{N-1}$ from a four-dimensional non-supersymmetric quiver gauge theory with
gauge group $SU(N)^{N_4N_6N_8}$. This theory is conformal in the large
$N_{4,6,8}$ limit.  We have given some evidence for the commutativity of the
following diagram, which summarizes the deconstruction:

\FIGURE{\input{diagram.epic} 
        \caption{Orbifold realization of the quiver gauge theory and 
          deconstruction of M-theory on $T^3 \times A_{N-1}$.}\label{fig3}}

The left hand side of the diagram shows the quiver gauge theory and its
realization in type IIB string theory as a stack of D3-branes placed at the
origin of an orbifold of the type $\CC^3/\ZZ_{N_4} \times \ZZ_{N_6} \times
\ZZ_{N_8}$. The right hand side represents the Higgs branch of the quiver
theory and its corresponding realization in M-theory.  Moving the D3-branes
away from the orbifold singularity corresponds to the Higgs branch of
the quiver theory. Exploiting string dualities, we map the geometry in the
vicinity of the D3-branes in type IIB string theory to a $T^3 \times A_{N-1}$
geometry in M-theory.  We conclude that the deconstructed Higgs branch
theory, in a particular strong coupling and large $N_{4,6,8}$ limit, is
equivalent to M-theory on $T^3 \times A_{N-1}$. This claim is further
supported by a dyonic spectrum in the quiver theory which corrsponds to 
wrapped M2-branes. We thus provide a completely field-theoretical definition of
M-theory on $T^3 \times A_{N-1}$.

The deconstruction of M-theory does not suffer from the problems of the
deconstruction of pure gravitational theories discussed in
\cite{Arkani-Hamed:2003vb, Arkani-Hamed:2002sp, Schwartz:2003vj, Jejjala,
  Jejjala:2003qg}.  In quiver {\em gauge} theories  the  cut-off for the
higher-dimensional behaviour of the theory can be taken to infinity.

However it would be of interest to find further field-theoretical evidence for
the equivalence.  For example, one would like to see how eleven-dimen\-sional
supergravity is realized in the model. The quiver model is formulated in terms
of local fields of an ordinary four-dimensional Yang-Mills theory. An open
question is the relation between these fields and the metric tensor or higher
spin fields in M-theory.  It would also be exciting to find a
field-theoretical argument for the quiver theory to describe eleven
dimensions besides the string-theoretical argument given in this paper.

The present 4d quiver theory provides a non-perturbative definition of
M-theory and might be an alternative to matrix models which describe the
discrete light-cone quantization (DLCQ) of M-theory \cite{Banks:1996vh,
  Susskind:1997cw}. It would be very interesting to find a relation between
both approaches. The matrix model for M-theory on $T^3 \times A_{N-1}$ is
given by a 4d $\N=2$ super Yang-Mills quiver theory with gauge group
$U(k)^N$~\cite{Ganor}. The parameter $k$ characterizes the discrete momentum
$P_-=k/R$ of states in the lightlike direction.  The Coulomb branch of this
model describes the gauge theory located at the singularity of the geometry.
The Higgs branch encaptures the physics of the spacetime away from the
singularity.  This matrix model has to be compared with the Higgs branch
theory of our model, which has unbroken gauge group $SU(N)$ and preserves 16
supercharges in the continuum limit.  The matrix model describes a sector of
M-theory with fixed momentum $P_-$. In contrast, our model is not
restricted on a particular sector of M-theory. Note however that the continuum
limit $a_{4,6,8}\rightarrow 0$ requires one to consider the quiver theory at
strong coupling, impeding perturbative access to M-theory.

\acknowledgments{
We wish to thank James Babington, Ralph Blumenhagen, Karl \mbox{Landsteiner},
\mbox{Esperanza} Lopez and Dieter L\"ust for useful discussions.  The
\mbox{authors} are particularly indebted to \mbox{Johanna} Erdmenger and
Zachary \mbox{Guralnik} for helpful comments.  The research of I.K.~is funded
by the DFG (Deutsche Forschungsgemeinschaft) within the Emmy Noether
programme, grant ER301/1-2.}

\end{document}